\documentclass[a4paper,11pt]{article}
\pdfoutput=1
\usepackage{jcappub}
\usepackage[T1]{fontenc} 
\usepackage{bm}

\usepackage{xcolor}

\newcommand{\f}{f(R)}
\newcommand{\fR}{f_R}
\newcommand{\fRR}{f_{RR}}

\newcommand{\fRz}{f_{R0}}

\newcommand{\compt}{\lambda_\mathrm{C}}
\newcommand{\comptbg}{\bar{\lambda}_\mathrm{C}}
\newcommand{\fRscr}{f_R^\mathrm{scr}}
\newcommand{\comptscr}{\lambda_\mathrm{C}^\mathrm{scr}}

\newcommand{\Mvir}{M_\mathrm{vir}}

\newcommand{\diff}[2]{\frac{\mathrm{d} #1}{\mathrm{d} #2}}
\newcommand{\diffn}[3]{\frac{\mathrm{d}^#3 #1}{\mathrm{d} #2^#3}}

\title{\boldmath Galactic Compton Wavelengths in $f(R)$ Screening Theories}

\author[a]{Bradley March,}
\author[a]{Clare Burrage}
\affiliation[a]{School of Physics and Astronomy, University of Nottingham, \\University Park, Nottingham NG7 2RD, UK}
\author[b]{and Aneesh P. Naik}
\affiliation[b]{Institute for Astronomy, University of Edinburgh, \\Royal Observatory, Blackford Hill, Edinburgh EH9 3HJ, UK}
\emailAdd{bradley.march@nottingham.ac.uk}
\emailAdd{clare.burrage@nottingham.ac.uk}
\emailAdd{aneesh.naik@roe.ac.uk}

\abstract{$f(R)$ theories of modified gravity may be compatible with current observations if the deviations from general relativity are sufficiently well screened in dense environments. In recent work \cite{PrevPaper} we have shown that approximations commonly used to assess whether galaxies are screened, or unscreened, fail to hold in observationally interesting parts of parameter space. One of the assumptions commonly made in these approximations, and more broadly in the study of $f(R)$ models, is that the mass of the scalar mode can be neglected inside a galaxy. In this work we demonstrate that this approximation may fail spectacularly and discuss the implications of this for tests of the theory.}

\begin{document}
\maketitle
\flushbottom

\section{Introduction}
\label{sec:intro}

In recent years, the continuing puzzle of the dark sector has motivated the study of scalar-tensor theories of gravity, in which the scalar field mediates gravitational-strength \textit{fifth forces} \cite{HuSawicki2007, Joyce:2014kja, Koyama:2015vza, Ishak:2018his, BurrageSakstein2018, Brax:2021wcv, CANTATA:2021ktz, Vardanyan:2023jkm}. For such a fifth force to evade detection in stringent Solar System tests of gravity, the theory must possess some \textit{screening mechanism}, such that the force is suppressed in environments resembling the Solar System (i.e., environments of high ambient density or deep gravitational potential). Various screening mechanisms have been proposed along these lines, for example, the Vainshtein and symmetron mechanisms \cite{HinterbichlerKhoury2010, Hinterbichler:2011ca, Vainshtein, BabichevVainshteinscreening}. In this work, we are concerned with the class of theories exhibiting \textit{chameleon} screening \cite{Khoury:2003aq, KhouryWeltman2004}, in which the mass of the scalar grows large in dense regions. In particular, we focus on the theory of Hu-Sawicki $f(R)$ gravity \cite{HuSawicki2007, Brax+2008}, which has received especially widespread attention in the literature.

Galactic scales present a particularly interesting area of study in $\f$ theories. Baker et al. \cite{Baker+2019} showed that galactic scales inhabit a `desert' in parameter space with comparatively few observable probes. In the years since that work, various studies have sought to explore this desert, with considerable success \cite{Hui+2009,JainVanderPlas2011,Vikram+2013,Vikram+2014,LombriserPenarrubia2015,Burrage+2017,Desmond+2018Warps,Desmond+2018Offsets,O'HareBurrage2018,Naik+2018,Naik+2019,Naik+2020Streams,Bartlett+2021,Pedersen:2023ina, Benisty:2023qcv}. Indeed, the strongest constraints to date on $f(R)$ gravity were obtained in recent works by Desmond and Ferreira \cite{DesmondFerreira2020}, and Landim et al \cite{Landim2024} in a search for chameleon-induced equivalence principle violations in a large sample of observed galaxies.

In parallel with these analyses of observable probes, there has been work aiming to improve our theoretical understanding of the interplay between chameleon fifth forces and galaxies. Because galaxies inhabit the deeply non-linear regime of cosmological structure formation, such an understanding can only be obtained by numerical computation, typically with various simplifying approximations. For example, recent work has produced cosmological galaxy formation simulations incorporating a quasi-static chameleon fifth force \cite{Arnold+2019}. Another example is our previous work \cite{PrevPaper} in which $f(R)$ solvers were applied to static galactic models to solve for the scalar field and accompanying fifth forces. For a review of the approaches and challenges to numerically modelling theories with screened fifth forces see ref.~\cite{llinares2018}.

When studying the behaviour of the scalar field within a galaxy, and the resulting screening of the fifth force, the various components of the galaxy (stars, gas and dark matter) are typically treated as continuous fluids of varying density. However, this approximation might not be a very good one in the case of the stellar component, or even the dark matter component if it is composed of compact objects and/or small-scale subhaloes. The goal of the present work is to test the approximation that the distribution of stars in a galaxy can be treated as a smooth density distribution. In particular, we calculate the Compton wavelength of the $f(R)$ scalar field across a series of typical galaxies, assuming a continuous stellar density profile, and compare this length scale with the typical stellar separation. To anticipate our results, we find that in the screened regions of galaxies, the Compton wavelength is often far smaller than the stellar separation, suggesting that a continuous fluid approximation is inadequate.

This work is structured as follows. In section \ref{sec:theory} we present the Hu-Sawicki $f(R)$ gravity theory, including expressions for the Compton wavelength. Section \ref{sec:numerical} describes our numerical solutions for the scalar field across our galactic model. Section \ref{sec:results} describes the results, before concluding remarks in section \ref{sec:conclusions}. Throughout this paper we work in natural units such that $8\pi G=c=\hbar=1$, except where units are explicitly stated. The code used to solve the field profiles and produce the figures contained within this paper is publicly available at \href{https://github.com/Bradley-March/scalar-compton-wavelength}{github.com/Bradley-March/scalar-compton-wavelength}.

\section{Theory}
\label{sec:theory}

\subsection{\texorpdfstring{$\f$}{f(R)} Gravity}
\label{ssec:f(R) gravity}

In $\f$ theories, modifications to GR are achieved by substituting the Ricci scalar, $R$, in the Einstein-Hilbert action with the generalised  $R+\f$. This allows the action to incorporate higher-order curvature terms, 
\begin{equation}
\label{eq:fR action}
    S = \frac{1}{2}\int d^4x \sqrt{-\tilde{g}}\left[R+\f\right] 
    + S_m\left[ \tilde{g}_{\mu\nu}, \psi^{SM}_i\right] \,,
\end{equation}
here $R=R(\tilde{g})$ represents the scalar-curvature of the Jordan frame metric. 

Such an action can be reformulated into the general scalar-tensor framework via the field redefinition \cite{Brax+2008}
\begin{equation}
\label{eq:fR field redefinition}
    \phi=-\frac{\ln(1+\fR)}{2\beta(\phi)} \,, \qquad 
    V(\phi) = \frac{R\fR - \f}{2(1+\fR)^2} \qquad
    \mathrm{and} \qquad
    \beta(\phi) = \frac{1}{\sqrt{6}} \,,
\end{equation}
where
\begin{equation}
    \fR \equiv \diff{\f}{R}
\end{equation}
serves as the scalar field in the theory. 

For certain functional forms of $\f$, the theory can display the chameleon screening mechanism \cite{Faulkner+2007f(R)ChameleonModels, Starobinsky2007f(R)ChameleonModels, NavarroVanAcoleyen2007f(R)ChameleonModels}. We adopt the well-known Hu-Sawicki model \cite{HuSawicki2007}. This model is characterised by the functional form
\begin{equation}
\label{eq:fR HS def}
    \f = - \frac{am^2}{1 + (R/m^2)^{-b}} \,,
\end{equation}
where $a$ and $b$ are positive dimensionless parameters, and $m$ has dimensions of inverse length. For simplicity, and consistency with the majority of literature, we set $b=1$. The remaining parameters, $a$ and $m$, can be related by ensuring that in the high curvature limit, $R\gg m^2$, gravity reverts to GR$+\Lambda$CDM, i.e. $\f\approx-2\Lambda$ where $\Lambda$ is the cosmological constant, giving
\begin{equation}
    am^2 = 2\Lambda \,.
\end{equation}

This allows the theory to be characterised by a single model parameter, which we choose to be the present-day background field value $\fRz$
\begin{equation}
\label{eq:fR constant parameters}
    a = -\frac{4\Omega_\Lambda^2}{(\Omega_m + 4 \Omega_\Lambda)^2} \frac{1}{\fRz} 
    \quad \mathrm{and} \quad
    m^2 = -\frac{3H_0^2 (\Omega_m + 4\Omega_\Lambda)^2}{2\Omega_\Lambda} \fRz \,,
\end{equation}
where $\Omega_m$ and $\Omega_\Lambda$ are the cosmological density parameters for matter and dark energy, respectively. Throughout this work, we use $\Omega_m=0.3$ and $\Omega_\Lambda=0.7$, except where specifically stated.

By extremising the action, eq.~\eqref{eq:fR action}, with respect to the metric we derive a set of modified Einstein field equations. Taking the trace and then applying the Newtonian limit, under the small field $|\fR| \ll 1$ and quasi-static approximations $|\nabla \fR| \gg \partial_t \fR$, we obtain the $\f$ equation of motion,
\begin{equation}
\label{eq:EoM}
    \nabla^2 \fR = \frac{1}{3}\left(\delta R - \delta\rho\right) \,,
\end{equation}
where $\delta\rho$ and $\delta R$ represent the density perturbation and curvature perturbation, respectively. Within the Hu-Sawicki model the curvature perturbation is defined as 
\begin{equation}
\label{eq:deltaR}
    \delta R = R_0 \left[\sqrt{\frac{\fRz}{\fR}} - 1\right] \,,
\end{equation}
where zero subscripts denote present-day background values. The acceleration due to the fifth force from our scalar field is 
\begin{equation}
\label{eq:fR 5th force}
    \bm{a}_5 = \frac{1}{2}\bm{\nabla} \fR \,,
\end{equation}
again assuming $|\fR| \ll 1$. 

An examination of the fifth force equation, eq.~\eqref{eq:fR 5th force}, reveals that the effect of the scalar field is minimised when the field is constant and its gradient is approximately zero. A region in which the field is constant is referred to as the screened region. An analytical expression for the field profile can be derived in this limit, by noting that a flat field profile also implies a vanishing Laplacian in the $\f$ equation of motion, eq.~\eqref{eq:EoM}. For the right hand side of this equation to vanish we require
\begin{equation}
\label{eq:screened field}
    \frac{\fRscr}{\fRz} = \left(1+\frac{\delta\rho}{R_0}\right)^{-2} \,,
\end{equation}
where $\fRscr$ represents the analytic approximation to the screened field value. This screened solution corresponds to the curvature tracing the minimum of the effective scalar potential. This method has been utilised in the literature, e.g. ref.~\cite{HuSawicki2007}. 

Outside of a few special simplified cases, for example the screened solution discussed in eq.~\eqref{eq:screened field}, it is not possible to solve the highly non-linear equation of motion for the $f_R$ field analytically. In later sections of this work, where we wish to understand the behaviour of the field in a galaxy,  we utilise a two-dimensional, cylindrically symmetric numerical solver, based on MG-GADGET \cite{Puchwein+2013MG-GADGET}, to solve a discretised version of the equation of motion, eq.~\eqref{eq:EoM}. For a given input $\fRz$ and $\delta\rho$, the code performs an iterative Gauss-Seidel method, to generate the field profile $\fR$. Section 3.2 of ref.~\cite{PrevPaper} provides a comprehensive description of the solver. 

\subsection{Compton Wavelength}
\label{ssec:Compton}

The Compton wavelength of our theory is given by
\begin{equation}
    \compt \equiv m_\phi^{-1} \,,
\end{equation}
where the scalar mass, $m_\phi$, is defined as
\begin{equation}
    m_\phi^2 \equiv \diffn{V(\phi)}{\phi}{2} \,.
\end{equation}
The scalar field, $\phi$, and potential, $V(\phi)$, are defined in eq.~\eqref{eq:fR field redefinition}. We compute the mass in the field picture ($\phi$, not $\fR$) because it is only in this picture that the scalar degree of freedom is canonically normalised, however, after applying the simplifications described later in this section both forms are equivalent to leading order.

The first derivative of the potential is 
\begin{align}
   \diff{V}{\phi} 
   = \diff{R}{\phi} \diff{V}{R} 
   = - \beta \frac{R + 2\f - R\fR}{(1+\fR)^2} \,, 
\end{align}
where we have used
\begin{equation}
    \diff{R}{\phi} = \left(\diff{\phi}{R}\right)^{-1} = -\frac{2\beta(1+\fR)}{\fRR} \,,
\end{equation}
given $\fRR\equiv\diff{\fR}{R}$.
The second derivative, resulting in the scalar mass, is
\begin{equation} 
\label{eq:scalar mass full}
    m_\phi^2=\diffn{V(\phi)}{\phi}{2} = \frac{2\beta^2}{(1+\fR)^2\fRR} \left[ (1+\fR)^2 - \fRR\left((3 - \fR)R + 4\f \right) \right] \,.
\end{equation}

We can simplify this scalar mass using the functional form of $\f$, eq.~\eqref{eq:fR HS def}, and its derivatives
\begin{align}
    \f &= -\frac{am^2R}{R+m^2} \,, \\
    \fR &= -\frac{am^4}{(R+m^2)^2} \,, \\
    \fRR &= \frac{2am^4}{(R+m^2)^3} \,.
\end{align}
In the high curvature regime, $R \gg m^2$, these expressions are well approximated as
\begin{align}
    \f &\approx -am^2 \,, \\
    \fR &\approx -\frac{am^4}{R^2} \,, \label{eq:field approx} \\ 
    \fRR &\approx \frac{2am^4}{R^3} \,. \label{eq:temp fRR approx}
\end{align}
Eliminating $R$ between equations~\eqref{eq:field approx} and  eq.~\eqref{eq:temp fRR approx} gives
\begin{equation}
    \fRR \approx \frac{2}{m^2}\sqrt{\frac{(-\fR)^3}{a}} \,.
\end{equation}

Substituting these expressions into the scalar mass, eq.~\eqref{eq:scalar mass full}, and assuming  $|\fR| \ll 1$, 
we obtain
\begin{equation} \label{eq:scalar mass temp}
    \frac{m_\phi^2}{2\beta^2 m^2} \approx 
    \frac{\sqrt{a}}{2}(-\fR)^{-3/2} 
    - 3\sqrt{a}(-\fR)^{-1/2} -a \,.
\end{equation}
Eq.~\eqref{eq:fR constant parameters} shows that $a \sim |\fRz|^{-1}$. Therefore, given that $|\fR| \ll 1$, the latter two terms in this equation are subdominant to the first, leaving us with the rather simple
\begin{equation} \label{eq:scalar mass simple}
    m_\phi^2 \approx \beta^2\sqrt{a}m^2(-\fR)^{-3/2} \,.
\end{equation}
This expression is equivalent to the previously obtained result, equation~(48) in ref.~\cite{HuSawicki2007}, albeit we have expanded the $\fRR$ term.

Using this simplified scalar mass, we determine the Compton wavelength of our theory to be
\begin{equation}
    \compt \approx \left(\frac{(-\fR)^3}{am^4\beta^4}\right)^{1/4} \,.
\end{equation}
By substituting the values for $\beta$ and the model parameters $a$ and $m$ (eqs.~\eqref{eq:fR field redefinition} and \eqref{eq:fR constant parameters} respectively), we arrive at the final expression
\begin{equation}
\label{eq:Compton wavelength}
    \compt \approx \frac{1}{H_0} \sqrt{\frac{2}{\Omega_m+4\Omega_\Lambda}} \left(\frac{\fR^3}{\fRz}\right)^{1/4} \,.
\end{equation}

In the cosmological background, $\fR\rightarrow\fRz$, the Compton wavelength reduces to
\begin{align}
\label{eq:background Compton wavelength}
    \comptbg &\approx \frac{1}{H_0} \sqrt{\frac{2|\fRz|}{\Omega_m+4\Omega_\Lambda}}.
\end{align}
In previous work, e.g. ref.~\cite{Cabre+2012}, this equation is often quoted as $\comptbg \approx  32\sqrt{\frac{|\fRz|}{10^{-4}}}\,\mathrm{Mpc}$, where the prefactor of 32 requires adopting cosmological parameters of $h=0.73$, $\Omega_m=0.24$ and $\Omega_\Lambda=0.76$.

\section{Galaxy Model}
\label{sec:numerical}

\subsection{Galactic Density Model}

To represent the galactic density profile we employ a two-component model, comprising a dark matter and stellar disc component. The dark matter is represented by the standard NFW profile \cite{NFW1996},
\begin{equation}
\label{eq:NFW density}
    \rho_\mathrm{DM}(r) = \frac{\rho_\mathrm{NFW}}{\dfrac{r}{r_\mathrm{NFW}} \left(1+\dfrac{r}{r_\mathrm{NFW}}\right)^2} \,,
\end{equation}
where $r$ is the spherical radial coordinate, and $\rho_\mathrm{NFW}$ and $r_\mathrm{NFW}$ are the characteristic density and length scales.
The stellar disc is represented by a double exponential profile 
\begin{equation}
\label{eq:stellar disc density}
    \rho_\mathrm{SD}(R,z) = \frac{\Sigma_\mathrm{disc}}{2z_\mathrm{disc}} e^{-R/R_\mathrm{disc}}e^{-|z|/z_\mathrm{disc}} \,,
\end{equation}
where $R$ represents the cylindrical radial coordinate, $z$ represents the coordinate perpendicular to the galactic disc, and $\Sigma_\mathrm{disc}$, $R_\mathrm{disc}$ and $z_\mathrm{disc}$ are the characteristic density and length scales.

In order to maintain the simplicity of our model and reduce the number of input parameters, we utilise several empirical relations to relate the various density profile scales to a single input parameter, the virial mass of the dark matter halo, $\Mvir$. The methodology employed to achieve this is outlined in appendix B of ref.~\cite{PrevPaper}. We cut off the central density singularity of the NFW profile by setting the density to be constant within 50\,pc. We also add a sharp outer cutoff at 2.2$R_\mathrm{vir}$ to address the logarithmically divergent mass \cite{Deason+2020}. Here the virial radius, $R_\mathrm{vir}$, is the radius at which the enclosed mass equals the virial mass, $M_\mathrm{vir}$.

\subsection{Separation Between Stars}

In section \ref{sec:results} we will discuss how the Compton wavelength of the $f_R$ field compares to the spatial separation between stars. Under the simplifying assumption of a single universal stellar mass $M_*$ and neglecting binary or multiple systems, the separation between stars is approximately given by
\begin{equation}
\label{eq:seperation}
    S = \left( \frac{\rho_{SD}}{M_*} \right)^{-\frac{1}{3}} \,.
\end{equation}
where $\rho_{SD}$ is the stellar density profile. 
In the remainder of this work we assume $M_*=M_\odot$ in this estimate. For a Milky-Way-like galaxy, we would estimate the typical stellar separation at a galactic radius of 8\,kpc is 2.5\,pc, which is of similar order to the Solar System-$\alpha$ Centauri separation of 1.3\,pc.

\section{Results}
\label{sec:results}

\subsection{Compton Wavelength in Galaxies}

\begin{figure}[htp]
    \centering
    \includegraphics[width=1\linewidth]{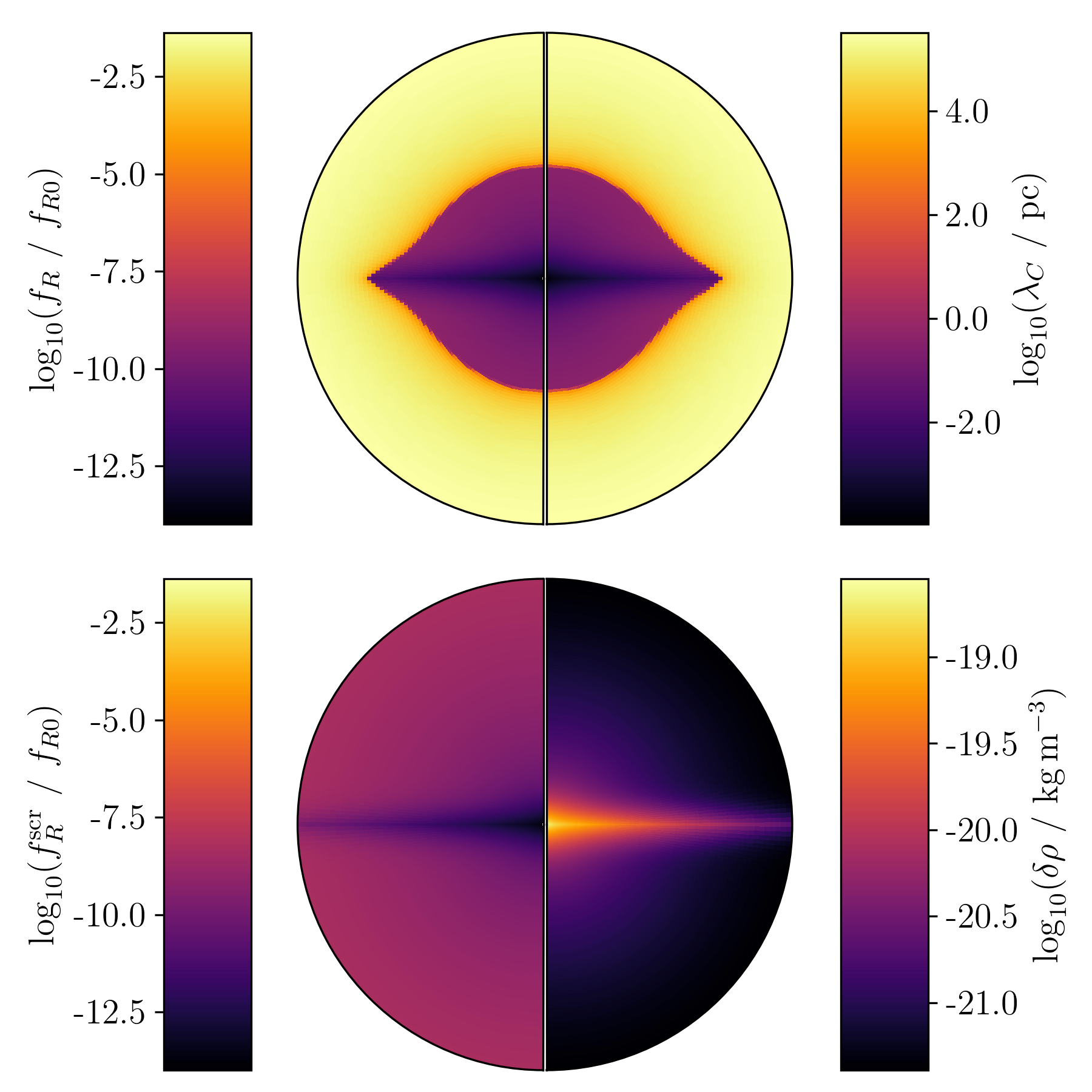}
    \caption{An example solution for a Milky-Way-like galaxy with $\Mvir = 1.5 \times 10^{12} M_\odot$, and $\fRz = -10^{-6}$, plotted to a maximum radius of 10\,kpc. Top left shows the field solution obtained from the numerical solver. Top right is the Compton wavelength, eq.~\eqref{eq:Compton wavelength}. Bottom left shows the screened field value, eq.~\eqref{eq:screened field}. Bottom right is the galactic density profile. This example illustrates a partially screened case, where the scalar field (and Compton wavelength) change dramatically at the screening surface, which is located within the stellar disc.}
    \label{fig1:exampleplot}
\end{figure}

Figure \ref{fig1:exampleplot} provides an example of the field profile, with a background value of $\fRz = -10^{-6}$, and related quantities that we can compute in a Milky-Way-like galaxy; specifically: the numerically calculated field profile, treating the stars as a continuous fluid, $f_R$, (top left); the Compton wavelength calculated from this field profile, eq.~\eqref{eq:Compton wavelength}, $\compt$, (top right); the analytic screened field value, $\fRscr$, eq.~\eqref{eq:screened field} (bottom left); and the galactic density profile, $\delta\rho$, eqs.~\eqref{eq:NFW density} and \eqref{eq:stellar disc density} (bottom right). The sharp transition in $\fR$ values shown in figure~\ref{fig1:exampleplot} corresponds to the position of the screening surface, $r_s$. Inside this surface fifth forces are suppressed and the field is small; outside, the field increases towards the background value of $\fRz$ and fifth forces are unsuppressed. For the parameters chosen in this figure, the galaxy is partially screened, with a screening surface located within the stellar disc. 

Figure \ref{fig1:exampleplot} also illustrates a number of relationships between parameters we have previously discussed. That $\compt \propto |\fR|^{3/4}$, eq.~\eqref{eq:Compton wavelength}, can be seen from the similarity of the profiles of the field (top left) and Compton wavelength (top right). We find that in the screened region the numerically calculated field (interior of the top left plot) is well approximated by the analytically derived screened field value (bottom left), i.e. $f_R(r<r_s) \approx \fRscr$. We also see the relationship of eq.~\eqref{eq:screened field} in the similarity between the density profile (bottom right) and the screened field (bottom left).

Considering the magnitude of the Compton wavelength in the top right panel of figure \ref{fig1:exampleplot}, we notice that outside the screened region the Compton wavelength is large, reaching values $\mathcal{O}(\mathrm{Mpc})$. However, inside the screened region the value of the Compton wavelength plummets to sub-parsec scales, significantly smaller than the typical separation of stars, eq.~\eqref{eq:seperation}.

\begin{figure}[htp]
    \centering
    \includegraphics[width=1\linewidth]{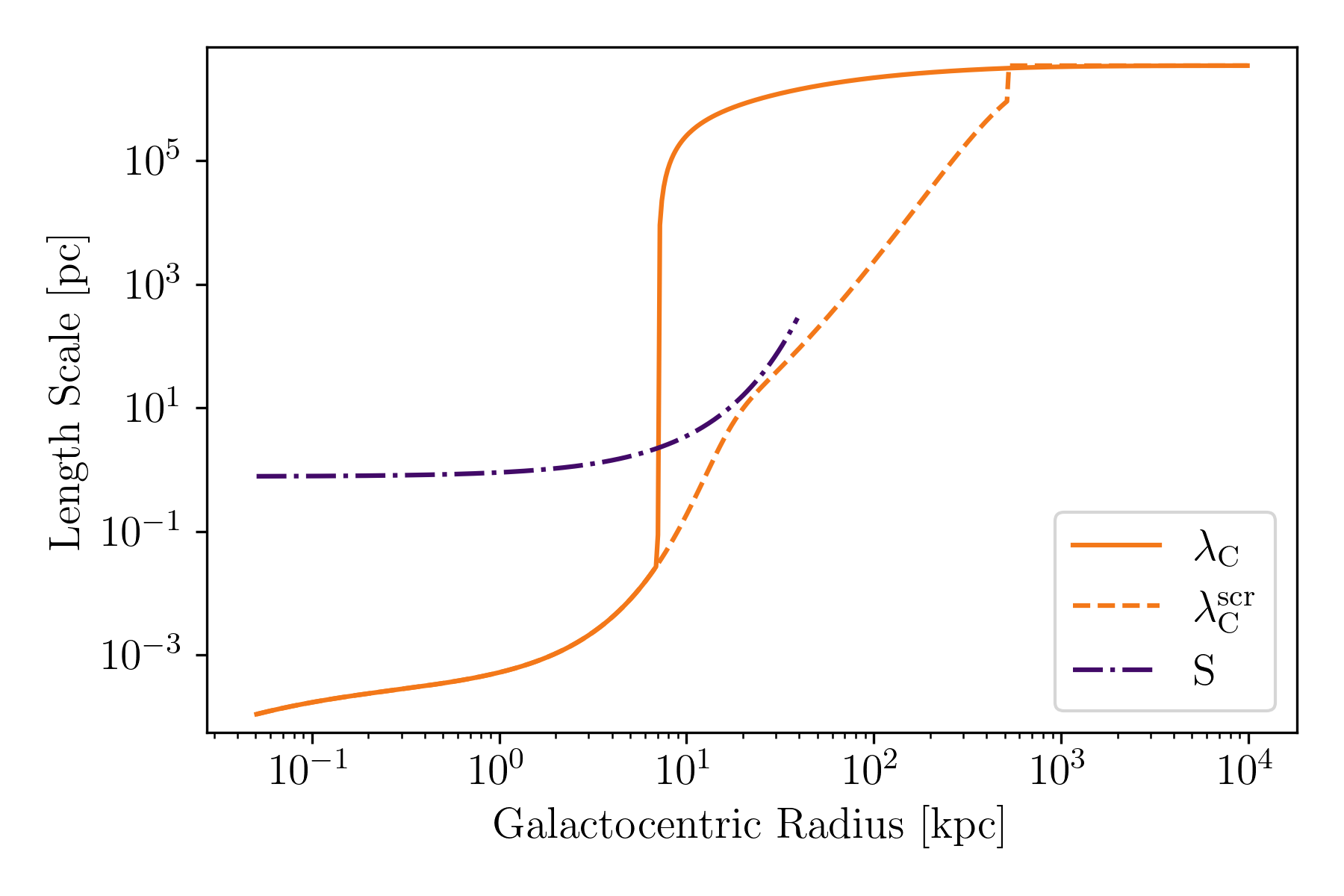}
    \caption{Comparison between average stellar separation (purple) and Compton wavelength (orange) as a function of galactocentric radius, in a Milky-Way-like galaxy with $\Mvir = 1.5 \times 10^{12} M_\odot$ and $\fRz = -10^{-6}$. Two curves are shown for the Compton wavelength: the numerical solution extracted from our solver (solid) and the analytic solution in the screened regime (eq.~\eqref{eq:screened field}; dashed). The curve representing the stellar separation has been truncated at the point where the stellar density is subdominant to the cosmic mean density. The figure shows that the Compton wavelength is orders of magnitude smaller(larger) than the stellar separation in the screened(unscreened) regime.} 
    \label{fig2:compt_vs_sep}
\end{figure}

We expand on this visualisation in figure \ref{fig2:compt_vs_sep} by comparing the lengths of the Compton wavelength and the stellar separation, eq.~\eqref{eq:seperation}, along the radial axis of the galactic disc, using the same parameters as in the previous figure. It is now evident that the Compton wavelength is orders of magnitude smaller than the stellar separation inside the screened portion of the galaxy, but rapidly grows to be much larger in the unscreened region of the galaxy.

\begin{figure}[htp]
    \centering
    \includegraphics[width=1\linewidth]{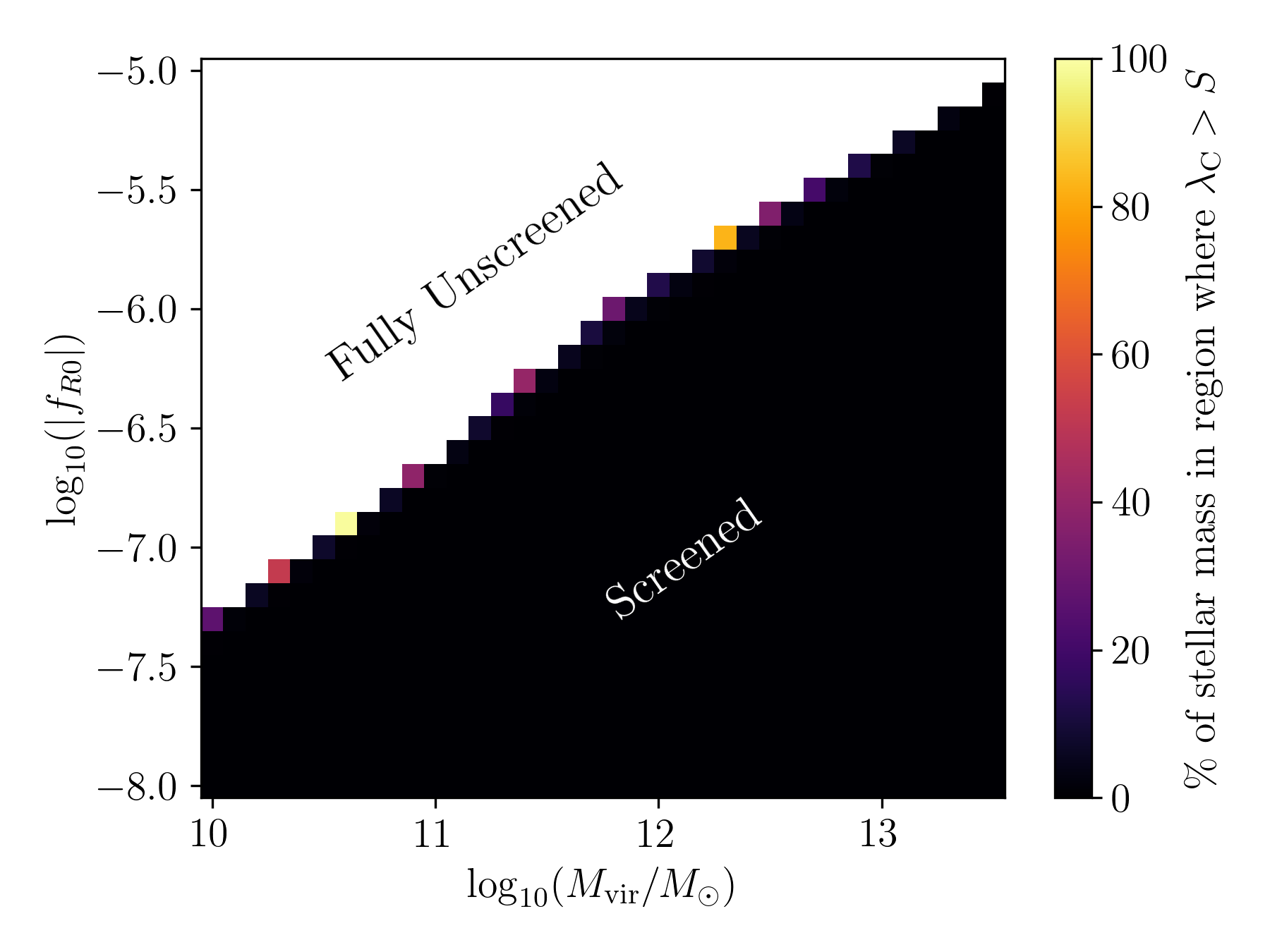}
    \caption{Proportion (in percent) of the stellar mass within the region where the Compton wavelength (calculated numerically, assuming a smooth stellar profile) is larger than the stellar separation, across a range of $f_{R0}$ and $\Mvir$ values. The top-left white region corresponds to fully unscreened solutions. This figure shows that the majority of galaxies that are not fully screened have stellar profiles with separations less than the Compton wavelength.} 
    \label{fig3:percent_valid}
\end{figure}

The parameters in figures \ref{fig1:exampleplot} and \ref{fig2:compt_vs_sep} were specifically chosen to simulate a galaxy with partial screening, where the screening radius lies within the stellar disc. In general, varying $\Mvir$ and $f_{R0}$, we find that most locations in parameter space result in a fully screened or fully unscreened stellar disc. Figure \ref{fig3:percent_valid} shows the proportion of stellar mass located in a region of the galaxy where $\compt > S$, for a range of $M_{\rm vir}$ and $f_{R0}$ parameters.  In the top left of the plot, where the galaxies are fully unscreened, we find $\fR \approx \fRz$ everywhere, and thus $\compt \approx \comptbg \gg S$. In the lower right region, the galaxies have fully screened stellar discs, resulting in $\compt = \comptscr \ll S$, as illustrated by the example in figure \ref{fig2:compt_vs_sep}. In the following section, we will further discuss the implications of the partially screened solutions, represented by the coloured (not black and white) points in figure \ref{fig3:percent_valid}.

\subsection{Scalar Field Values in the Interstellar Medium}

At a given point in space, the scalar field is sourced primarily by mass within a few Compton wavelengths; the contributions of more distant matter are exponentially suppressed. Where the Compton wavelength is larger than the typical separation between objects, the field profile evolves as if the density profile of the individual sources had been smoothed, with a smoothing scale of order the Compton wavelength. When the Compton wavelength is less than the typical separation between objects, there is no smoothing effect, and the field responds to the objects individually. When studying the behaviour of the $\f$ field within a galaxy it is common to treat the stellar component as a fluid with a smooth density profile (for example refs.~\cite{Arnold+2019,DesmondFerreira2020}). As we have seen, this approximation is not valid in much of the interesting parameter space.  In this section we consider how the field might behave in the absence of the smoothing of the stellar density profile. 

In the small Compton wavelength regime, it becomes invalid to consider the stellar density profile as a continuous distribution. As a result, we expect the scalar field to take differing field values near to stars, and deep in the interstellar medium. This is impossible to study analytically and will be challenging to solve numerically given the dynamical range of the problem. However, we can suggest some behaviour of the field in this regime. 
We expect that, in between the stars, the field will relax to the dark-matter-only solution, $\fR^\mathrm{DM}$, with pockets around the now-discrete stars where the field becomes orders of magnitude smaller. Therefore, except in the neighbourhood of the stars, the field evolves as if the stellar component of the galaxy was absent. 

There is also a rare, more complex case where, if we include the smoothed stellar mass in the galactic density profile, the Compton wavelength is smaller than the stellar separation, but removing the stellar contribution from the density profile, the dark-matter-only solution results in a Compton wavelength larger than the stellar separation (i.e. $\compt^\mathrm{DM+SD}<S<\compt^\mathrm{DM}$). In this highly non-linear regime, we suggest that the field in the interstellar medium relaxes to larger values causing an increase in the Compton wavelength until the Compton wavelength becomes large enough to be comparable to the stellar separation. A similar situation to this, although on very different scales, was considered in ref.~\cite{Brax:2013cfa}. In this situation, we suggest that the typical field value will be that which gives a Compton wavelength of order the stellar separation. However, further numerical study would be needed to understand this case in detail.

\begin{figure}[htp]
    \centering
    \includegraphics[width=1\linewidth]{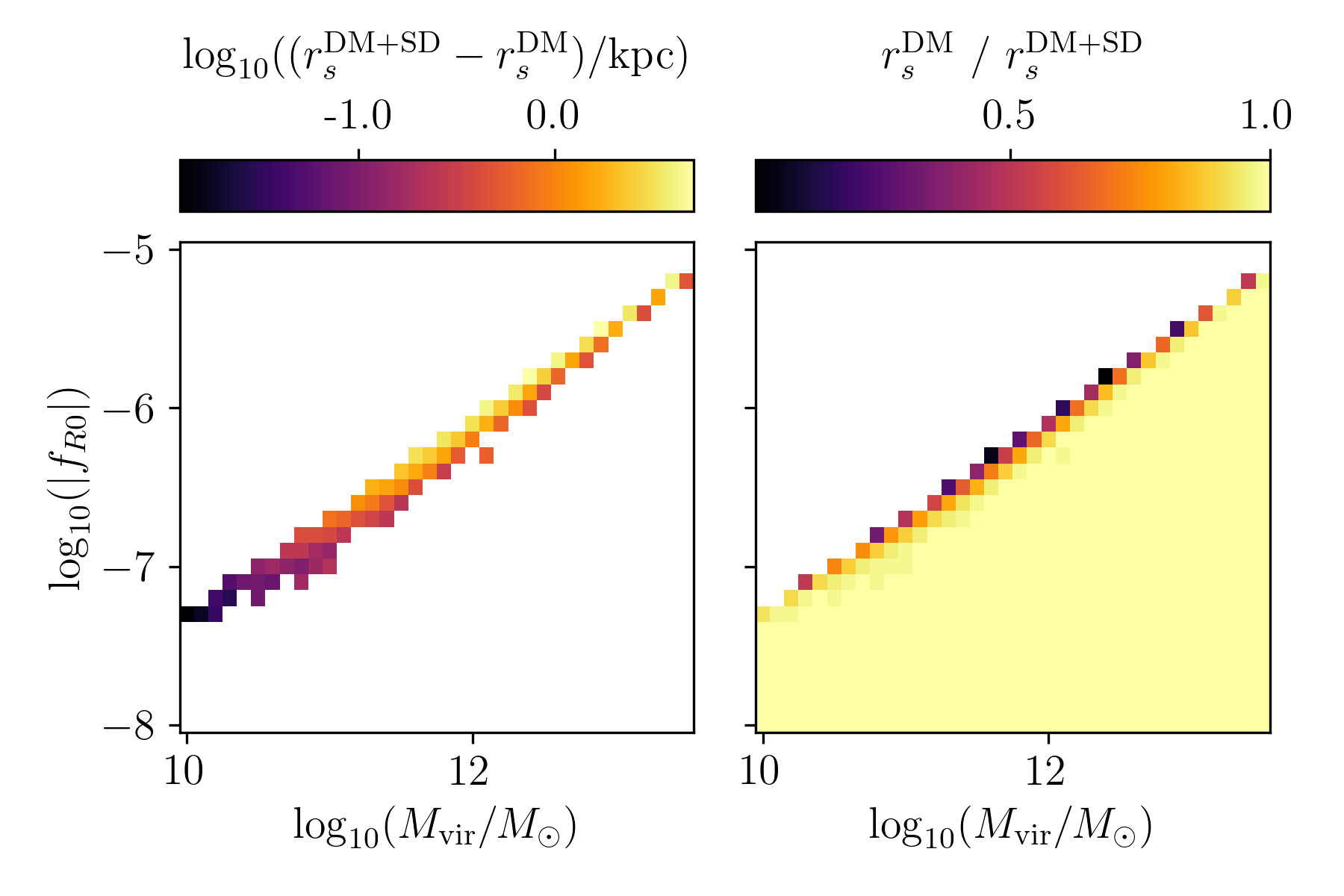}
    \caption{Comparisons of the screening radius assuming no stellar contribution to the galactic density ($r_s^\mathrm{DM}$) and assuming a smooth stellar density profile ($r_s^\mathrm{DM+SD}$). Left shows the absolute distance between the two $r_s$ solutions, and right shows the ratio between them. In both cases, the upper-left white region corresponds to fully unscreened solutions. In the lower-right regions, the screening radii are larger than the extent of the stellar profile and thus are unchanged by ignoring the stellar contribution. In the right plot these correspond to ratios of 1, in the left we have masked these values. Assuming no stellar contribution has the largest impact on solutions where the screening radius is within the stellar disc, corresponding to the thin diagonal highlighted in the left plot, or the ratios with a value less than one in the right plot.} 
    \label{fig4:rs_ratio}
\end{figure}

If it is the case that treating the stellar population as a discrete distribution would result in the field decreasing from $\fR^\mathrm{DM+SD}$ to $\fR^\mathrm{DM}$, then there would be a knock-on effect on the position of the screening surface, $r_s$. In figure \ref{fig4:rs_ratio} we compare the screening radii derived in the two cases: by assuming the stellar profile is a continuous distribution that contributes to the field profile, $\fR^\mathrm{DM+SD}$; and by neglecting the contribution of the stellar density profile, i.e. a dark matter only solution, $\fR^\mathrm{DM}$. In both cases we calculate the screening radius using the curvature-to-density parameter derived in our previous work \cite{PrevPaper}. 

Figure \ref{fig4:rs_ratio} shows that if $r_s$ is situated within the stellar disc, and $\compt < S$, then the decrease in the field values would act to shift $r_s$ towards the centre of the galaxy. Overall this would lead to less of the galaxy being screened, in this narrow diagonal band of parameter space. Although this only affects a small part of the parameter space, it remains relevant for current constraints, since many contemporary tests rely on partially screened stellar populations as observable tracers of equivalence principle violations.

\subsection{Fifth Forces Around Stars}

Compton wavelengths that are shorter than the interstellar separation, mean that there can be significant variation in the value of the field in the space between the stars.  Here we provide an estimate of whether this can lead to observable fifth forces in the Solar System. 

We approximate the Sun as a sphere of constant density, embedded in a background in which the field is screened, $\fR=\fRscr$. The star is screened as long as $|\fR|\lesssim \Phi_N(R_*)$ where $\Phi_N(R_*)$ is the Newtonian potential evaluated at the surface of the star. As $|\fRscr|\ll|\fRz|$, this is a weaker condition than that normally applied, which states that a star is screened if $|\fRz|\lesssim \Phi_N(R_*)$. At the position of the Sun, $\fRscr \sim 4 \times 10^{-11} \fRz$, and $\compt \sim 10^{18} \fRz^{1/2}\,\mathrm{m}$. The fifth force, in this approximation of the Solar System, is 
\begin{equation}
    \frac{F_5}{F_N} \approx \left(\frac{|\fR|}{5\times10^{-6}}\right) \left(\frac{R_\odot}{r}\right)^2 e^{-r/\lambda_C} \,,
\end{equation}
which is extremely small, and well beyond the sensitivity of current probes.

\section{Conclusions}
\label{sec:conclusions}

Scalar fields react to matter on scales equivalent to the Compton wavelength of the field. When modelling populations of discrete objects, it is common practice to treat the profiles as continuous distributions, assuming that many discrete objects are contained within a Compton wavelength, acting to smooth out the scalar field profile. 

In the context of chameleon $\f$ gravity in galactic environments, surveys are beginning to probe the regime where the screening radius may lie within the galactic disc and where the Compton wavelength can become significantly smaller than the typical separation of stars within the galaxy. As such, it is important to exercise caution when assuming a continuous stellar density profile. 

A proper treatment of the stars as individual discrete objects may result in a heterogeneous field profile, in which the mass of the stars is inconsequential to the field in the interstellar medium, allowing the interstellar field to relax to a larger field value permitted by the dark matter density only solution. In regions closer to the high-density stars, the scalar field will become further screened, adopting smaller, more screened field values. Such a situation could cause galaxies with partially screened stellar discs to become less screened, as neglecting the stellar mass profile pushes the screening surfaces towards the centre of the galaxy. In this case, there is the possibility that modified gravity tracers that rely on the galaxy being unscreened could become observable once more.

In conclusion, we should always be careful to check that the assumptions we are making actually hold in the environments we study, and as the accuracy of observations improves we should also improve the accuracy of our theoretical predictions.

\section*{Acknowledgements}
B.M. is supported by an STFC studentship. C.B. is supported by the STFC under grant
ST/T000732/1. A.P.N. is supported by an Early Career Fellowship from the Leverhulme Trust.
For the purpose of open access, the authors have applied a Creative Commons Attribution
(CC BY) licence to any Author Accepted Manuscript version arising.

\appendix

\bibliographystyle{JHEP}
\bibliography{library}

\end{document}